\documentclass[a4paper,11pt]{article}

\usepackage{pos}

\title{Cosmological studies with VLBI}

\author*[a]{Cristiana Spingola}

\affiliation[a]{INAF $-$ Istituto di Radioastronomia,\\
 Via Gobetti 101, I$-$40129, Bologna, Italy}

\emailAdd{spingola@ira.inaf.it}

\abstract{Current cosmological controversies can be solved if a sufficient level of precision is achieved by observations. Future surveys with the next generation of telescopes will offer significantly improved depth and angular resolution with respect to existing observations, opening the so-called “era of precision cosmology”.
But, that era can be considered already started at the radio wavelengths with Very Long Baseline Interferometry (VLBI). In this paper, we give an overview on how VLBI is contributing to some open questions in contemporary cosmology by reaching simultaneously the largest distances and the smallest scales.}

\FullConference{%
  15th European VLBI Network Mini-Symposium and Users' Meeting (EVN2022)\\
  11-15 July 2022\\
  University College Cork, Ireland
}

\begin{document} 
\maketitle


\section{Introduction}

In the current standard cosmological framework $\Lambda$CDM (Dark Energy Cold Dark Matter), simple gas (mostly H and He) evolves into highly complex galaxies, which are gravitationally bound systems made of metal enriched gas, stars, dust, supermassive black holes, that can become active and have a strong effect on the surrounding medium. All of these components are believed to be embedded in a dark matter halo \cite{Cimatti}. The cosmological model expects that the growth of these structures is driven by the dark matter component and happens in an expanding Universe, as known for almost a century \cite{Hubble1929} but still controversial \cite{Riess2020}. We are in an era where we can approach cosmology from an observational rather than a theoretical standpoint. The $\Lambda$CDM model has been successful at explaining the observations of the large-scale structure of the Universe, but it presents some issues on the small (galactic and sub-galactic) scales \cite{Bullock2017}. To clearly assess if such small-scale problems affect \textsl{all} galaxies, it is mandatory to robustly characterize also high redshift (high-z) galaxies. To do so, a primary requirement is high angular resolution and deep observations.

Current VLBI arrays are ideal to perform such tests, as they can provide milliarcsecond (mas) and sub-mas angular resolution (at high frequency) and $\mu$Jy beam$^{-1}$ sensitivity observations. However, if we want to use VLBI arrays, the selected targets must emit radiation at the radio wavelengths. Among the possible sources to investigate the small-scales of high-z galaxies there are active galactic nuclei (AGN). In particular, those AGN that shows collimated jets ("jetted AGN") can reach extremely large luminosities because of the effect of Doppler boosting \cite{Padovani2017}. Therefore, their radio emission can be detected up to the highest redshifts, making them ideal targets for studying the most distant Universe. Since jetted AGN can be cosmologically distant sources, some of them could be \textsl{strongly gravitationally lensed} (see \cite{Treu2010} for a review on strong lensing). The magnification and amplification due to the presence of a foreground (lensing) massive and concentrated object can make possible the detection also of the faintest jetted AGN in the form of multiple  and distorted images. Thus, strong lensing can provide an unbiased view on the jetted AGN population across the cosmic time.  Both lensed and "unlensed" jetted AGN give valuable observational constraints that can directly test  the standard model.


In this paper, we describe the contribution of VLBI to some current open issues in cosmology related to AGN and their surrounding medium (Sec.~\ref{Sec:2}), dark matter (Sec.~\ref{Sec:3}) and dark energy (Sec.~\ref{Sec:4}). Finally, we conclude by discussing the role of current and future wide-field VLBI surveys for observational cosmology (Sec.~\ref{Sec:5}).

\section{AGN and molecular gas at high redshift}\label{Sec:2}

How supermassive black holes (SMBHs) form and evolve is still unclear. The presence of SMBHs of masses of $10^{8-9}$ M$_{\odot}$ when the Universe was less than a Gyr old \cite{Banados2018, Belladitta2022} supports models of high mass BH seeds and super- (or even hyper-) Eddington accretion events. Some models assume an initial phase dominated by BH-BH mergers lasting up to $z\sim6$. After this epoch, the accretion of the resulting SMBHs is believed to be driven by inflowing of their surrounding medium (e. g., \cite{Piana2021}). From an observational point of view, jetted AGN have been found in gas-rich environments, which supports some of these models and at the same time suggests that mergers could be the triggering of radio-loudness in AGN \cite{Chiaberge2015, Gao2020}. 
Therefore, observations of the molecular gas surrounding jetted AGN can be a powerful method to test both SMBHs formation models and the activation of the radio-loudness across the cosmic time.

Very recently, the James Webb Space Telescope (JWST) has opened a new frontier for the study of high redshift galaxies, with a large number of candidates \textsl{during} the reionization epoch (at photometric redshifts $z>7$, e.g. \cite{Whitler2023} and references therein). The observations with JWST are suggesting even older and more massive SMBHs than expected, which require complex super-Eddington growth \cite{Pacucci2022, Volonteri2023}. 
VLBI observations of jetted AGN are currently reported up to $z=6.82$ \cite{Momjian2021}, finding jets extended from a few to several hundreds of parsecs \cite{Momjian2008, Frey2010, Frey2011, Frey2013, Cao2014, Gabanyi2015, Zhang2017,Cao2017, Wang2017, Momjian2018, Frey2018, Perger2019, Spingola2020, An2020, Zhang2021, Gabanyi2021, Perger2021, Zhang2022, Krezinger2022}. 
As for the molecular gas, at high-z detections of serveral J transitions of the CO and C$^+$ emission lines have been reported in the radio/mm bands \cite{Walter2014, Aravena2016a, Aravena2016b,  Decarli2016, Riechers2019, Spingola2020_gas, Bischetti2021}, which revealed the presence of massive molecular gas reservoirs around SMBHs, in agreement with the predictions from hydrodynamical simulations \cite[e.g.,][]{DiMatteo2008,Dave2020}. 
Moreover, detailed analysis of the profile of these emission line showed evidence for on-going mergers in gas-rich radio-loud AGN at $z>6$ \cite{Khusanova2022}. However, these results have been obtained for the brightest high-z sources and these sources were only partially spatially resolved. In this context, the magnification due to strong lensing is really powerful for detecting the low surface brightness emission of the low-J transitions of CO at high redshift. This boost in spatial resolution and sensitivity revealed both mergers (complex velocity fields) but also molecular gas disks (ordered rotation) in jetted AGN at high redshift \cite{Riechers2008a, Riechers2008b, Riechers2011,Deane2013, Paraficz2018, Spingola2020_gas}.

As an intermediate stage of the merger process, we expect to observe binary (or multiple) SMBH systems \cite{Begelman1980, DiMatteo2005, Volonteri2021}.  The interesting multiple SMBHs systems for the future pulsar timing arrays and the Laser Interferometer Space Antenna are those at \textsl{small separation}, as they will be the emitters of gravitational waves in the $\mu$Hz--nHz regime \citep{BurkeSpolaor2019, Colpi2019}. The fraction of the active pairs of SMBHs (binary/dual AGN) is very low in the local Universe, but it increases up to a few percent at high redshift \citep{DeRosa2019}. Hence, ideally, the search for close pairs of SMBHs should be performed at $z>1$. Only VLBI can spatially resolve such systems on parsec scales even at large distances. Indeed, using VLBI observations several pairs of AGN have been found so far \citep{Rodriguez2006, Bondi2010, Burke-Spolaor2011, An2013, Deane2014, Gabanyi2016}.

The combination of strong lensing and VLBI has been proven to be a powerful method to unveil dual AGN candidates at high redshift and small separation. Three cases have been reported so far:  MG~B2016+112 at $z=3.273$ \citep{Spingola2019_mg2016, Schwartz2021}, PS J1721+8842 at $z=2.37$ \citep{Mangat2021} and PKS~1830-211 at $z=2.5$ \citep{Nair2005}. As first suggested by \citep{Spingola2019_mg2016}, the detection of even a single strongly lensed binary AGN implies a larger population of AGN pairs in the early Universe than first thought. However, such estimate is highly limited by the very small number statistics, and these lensed (and unlensed) candidates should also be confirmed. 
Nevertheless, VLBI provides currently the only way to find the closest AGN pairs, hence probing the bound systems and the final merging stages at the largest distances (as summarized in Fig. 7 of \citep{Spingola2022}).

\section{Dark matter distribution at high-z}\label{Sec:3}


Among the current small-scale issues of the $\Lambda$CDM model, the most persistent ones are the so-called \textsl{missing satellite} and the \textsl{cusp-core} problems \cite{Bullock2017, Delpopolo2021}. Both these observational-theoretical conflicts could be solved by questioning the cold dark matter paradigm and assuming an alternative ("warm") dark matter particle \citep{Lovell2012}. 
Gravitational lensing provides a direct way to both find the expected satellites orbiting around massive galaxies and infer the inner mass density profile to the the cusp-core problem. In particular, the strong lensing systems showing many extended images in radial and tangential directions are ideal for such investigations. One can detect the sub-halos associated with a lensing galaxy or along the line of sight by means of "anomalies" in the flux, position or time delay in the lensed images. 
Such "anomalies" arise because the presence of a sub-halo that locally changes the magnification of that image, hence its properties (in particular flux and position) will be different from what expected for a smooth mass density distribution (i.e. without sub-halos). For instance, \citep{Hartley2019} recently detected mas-level positional anomalies in the lensed images positions inferred by VLBI observations in HS 0810+2554. This system shows eight extended lensed images of a background AGN with intrinsic integrated flux density of only 880 nJy.
Given the angular scale of these astrometric perturbations, they could be due to a population of low-mass (10$^{5-7}$ M$_{\odot}$) dark matter dominated sub-halos. 
Milliarcsecond-level astrometric anomalies were found also using global-VLBI observations of the lensing system MG J0751+2716, which shows compact and extended lensed images forming giant gravitational arcs \citep{Spingola2018}. Therefore, also for this system it is possible that the observed astrometric anomalies are due to low-mass sub-halos (see also \citep{Biggs2004, McKean2007, More2009, Hsueh2016, Hsueh2017, Stacey2020}). The inner density profile of MG J0751+2716 was inferred at high precision, thanks to the many constraints from the VLBI-detected lensed images. The inner mass density profile has been found to be steeper than isothermal: such steepening could be caused by the dissipative process due to \textsl{in situ} star formation, which leads to an an increased stellar content in the central regions of the lensing galaxy. Therefore, it is possible that the complex processes involving \textsl{baryonic} matter might be responsible for the current tensions regarding the small scales of galaxies. Only VLBI observations of extended gravitational arcs can test such complex models, as demonstrated by \citep{Powell2021, Powell2022} for testing the origin of the astrometric anomalies in MG J0751+2716.

Despite the advantages of using gravitational lensing and VLBI to study astrometric anomalies at mas and sub-mas level, we are currently limited by the small number of VLBI-detected systems. More strongly lensed jetted AGN observed with VLBI are needed to provide a conclusive interpretation on these dark matter studies.

\section{Cosmological parameters from jetted AGN observed with VLBI}
\label{Sec:4}


The research field of using VLBI observations of AGN to infer cosmological parameters (such as $q_0$ and $H_0$) has been pioneered by \citep{Gurvits1993, Kellerman1993}. They proposed a characteristic angular size $\theta$ for compact jetted AGN to build a cosmic distance ladder. For a cosmological rod of intrinsic length $l_m$, the $\theta$-z relation can be written as $\theta(z) = l_m/ D_A(z)$, where $D_A$ is the angular diameter distance, which depends on the cosmological parameters. This method was applied for the first time using $\sim340$ VLBI observations of AGN with known redshift by \cite{Gurvits1994}. With these observations they built a $\theta-z$ plane up to $z\simeq4$ and could test different cosmological models (namely, different values of $q_0$), despite the large uncertainty given by the limited number of sources.  Also, early work of \cite{Homan2000} proposed to measure the distance to jetted AGN by combining the observed proper motion, Doppler factor derived from equipartition between the energy density in magnetic and particles fields, and aspect ratio of the (superluminal) jet component. Recently \cite{Cao2019_curvature} proposed to use VLBI-detected AGN also to measure the cosmic curvature ($\Omega_k$), finding no significant deviations from a flat Universe.
The caveat about these methods is whether it is possible to use jetted AGN as standard rods. For this reason, a lot of research has been done to find the best jetted AGN "category" or some of their components for this cosmological application (see \citep{Cao2017_rulers} and references therein).

Another methodology to test cosmological models is to measure jet proper motions ($\mu$) across the cosmic time, as $\mu$ depends on the cosmological parameters \citep{Cohen1988, Kellerman2004, Jorstad2017}. The difficulty of this method is to obtain measurements of proper motions at high-z, as such measurements require long time baselines and high signal-to-noise ratios. Nevertheless, such measurements have been obtained up to $z\simeq 5.5$ (e.g., \citep{Frey2015, Perger2018, An2020, Zhang2022} and references therein). Recently, a reverse application of the $\mu$-z plot has been proposed, namely for deriving the redshift of a radio-loud AGN by adopting the standard cosmological model \citep{An2020_mu_z}. This method will be particularly important for wide-field surveys, where there are AGN candidates lacking spectroscopic redshift.

A method to estimate $H_0$ is based on the variability of jetted AGN \cite{Wiik2001} and has been recently used by
\cite{Hodgson2020} on the Boston University blazar monitoring  programme \cite{Jorstad2016}. The approach is based on light travel-time ("causality") arguments, and can be extended from low- to high-z AGN. The angular diameter distance $D_A$ can be also expressed as $D_A(z) = c \Delta t / \theta_{\rm VLBI} (1+z)$, where $\Delta t$ is the variability  timescale and $\theta_{\rm VLBI}$ is the size of size of the variable region. By applying this method on the low-z AGN 3C~84, they infer a value for $H_0$ with a similar precision of other "late Universe methods" applied to samples of sources (summarized in \citep{Verde2019}), demonstrating  that this approach is a competitive cosmological probe. Recently, \cite{Hodgson2023} revised the most important issue of this methodology, which is the determination of the Doppler factor, which becomes more difficult to obtain for high-z jetted AGN. The authors propose a method to derive the Doppler factor that is independent from the typical equipartition and inverse Compton brightness temperature limits \cite{Readhead1994}.

Finally, gravitational time delays observed in strong lensing systems provide a one-step measurement for $H_0$ \citep{Refsdal1964}, because the time-delay distance is inversely proportional to $H_0$ \cite{TreuMarshall2016}. Even if time delays between the multiple images can be estimated also using a spatially unresolved light curve (e.g., \cite{Barnacka2015}), at the radio wavelengths i) VLBI can easily resolve the lensed images (typically separated by arcseconds), ii) AGN show intrinsic time variability on several time scales and iii) there is not the additional challenge of microlensing, as the (background) jetted AGN are typically extended several parsecs in size. Despite the powerful combination of VLBI and gravitational time delays, there has not been any time delay measured using VLBI yet. This is mostly due to the high cadence of the observing epochs required and the relatively short periods for such monitoring programs available  (e.g., \citep{Spingola2016}).  Instead, monitoring programs with the Very Large Array have been proven to be quite effective for estimating gravitational time delays also by innovatively using  polarization \citep{Biggs2018a, Biggs2018b, Biggs2021}.

\section{Wide-field VLBI for cosmological probes and conclusions}\label{Sec:5}

Wide-field VLBI is now possible routinely. In the last $\sim10$ years, impressive progress has been done to make this possible, and in particular the development of multi-phase centre correlation (i.e., the entire primary beam of VLBI experiment can be imaged \cite{Deller2011, Morgan2011}) and multi-source self-calibration \cite{Radcliffe2016}. The progress in wide-field VLBI is timely, as there is the need to complement current (and future) surveys with radio observations at sub-arcsec angular resolution. At the same time, the exquisite sensitivity that wide-field VLBI surveys can reach now (e.g., $1.5\,\mu$Jy beam$^{-1}$ in the \textsl{e}MERGE survey \cite{Muxlow2020}) enables us to study at high image fidelity the faintest AGN population, which only the lensing magnification revealed so far \cite{Jackson2011, Hartley2019, Badole2020}.
Hence, wide-field VLBI surveys can detect the different radio sources that are most suitable for each of the cosmological applications described in this paper. For instance, the mJIVE-20 survey at 1.4 GHz \citep{Deller2014} has been used to search for radio-loud AGN in gas-rich mergers \cite{Shabala2017}, to find strongly lensed jetted AGN at small angular separation \citep{Spingola2019_mjive} and for the hunt for AGN pairs \citep{Burke-Spolaor2015}. The same wide-field VLBI survey has been also used for developing tools to identify and characterize VLBI sources with deep learning \cite{Rezaei2022}. Such tools together with pipelines for calibrating large VLBI datasets (e.g., \cite{Janssen2019}) are crucial for current and future wide-field VLBI surveys.


In summary, VLBI can provide precise and independent cosmological probes. In this paper, we focused on how VLBI observations of AGN (lensed and unlensed) can test the current cosmological model. Future surveys with the Square Kilometer Array (SKA) and the follow-ups with SKA-VLBI will offer significantly improved sensitivity. In the future,  we will not be limited by the small number statistics anymore, but we will also need synergy with multi-wavelength instruments to more robustly provide competitive precision on the estimate of the cosmological parameters.


\bibliographystyle{JHEP}
\bibliography{refs}

\end{document}